\documentclass[aps,prd,twocolumn,superscriptaddress,preprintnumbers,floatfix,nofootinbib,notitlepage,showkeys]{revtex4-1}

\usepackage[utf8x]{inputenc}
\usepackage{cancel}

\usepackage{graphicx}
\usepackage{hyperref}
\usepackage{latexsym}
\usepackage{amsmath}
\usepackage{amssymb}
\usepackage{bbm}

\usepackage{ulem}
\usepackage{pdfsync}
\usepackage{epsfig}
\usepackage{epstopdf}
\usepackage{subfigure}
\usepackage{color}
\usepackage{comment}
\usepackage{slashed}

\def\beq{\begin{equation}}
\def\eeq{\end{equation}}
\def\baq{\begin{eqnarray}}
\def\eaq{\end{eqnarray}}

\newcommand{\be}{\begin{equation}} % only untightened
\newcommand{\ee}{\end{equation}}
\newcommand{\bea}{\begin{eqnarray}} % only untightened
\newcommand{\eea}{\end{eqnarray}}

\newcommand{\bmp}{\noindent\begin{minipage}{16cm}}
\newcommand{\emp}{\end{minipage}\vskip 7mm} % 7mm untightened
\def\lsim{\mathrel{\raise.3ex\hbox{$<$\kern-.75em\lower1ex\hbox{$\sim$}}}}
\def\gsim{\mathrel{\raise.3ex\hbox{$>$\kern-.75em\lower1ex\hbox{$\sim$}}}}

\newcommand{\intron}[1]{}%{#1}

\begin{document}

\title{Observational Constraints on Decoupled Hidden Sectors}

\author{Matti Heikinheimo}
\email{matti.heikinheimo@helsinki.fi}
\affiliation{Department of Physics, University of Helsinki \\
                      P.O.~Box 64, FI-00014, Helsinki, Finland}
\affiliation{Helsinki Institute of Physics, \\
                      P.O.~Box 64, FI-00014, Helsinki, Finland}  
\author{Tommi Tenkanen}
\email{tommi.tenkanen@helsinki.fi}
\affiliation{Department of Physics, University of Helsinki \\
                      P.O.~Box 64, FI-00014, Helsinki, Finland}
\affiliation{Helsinki Institute of Physics, \\
                      P.O.~Box 64, FI-00014, Helsinki, Finland}  
\author{Kimmo Tuominen}
\email{kimmo.i.tuominen@helsinki.fi}
\affiliation{Department of Physics, University of Helsinki \\
                      P.O.~Box 64, FI-00014, Helsinki, Finland}
\affiliation{Helsinki Institute of Physics, \\
                      P.O.~Box 64, FI-00014, Helsinki, Finland}  
\author{Ville Vaskonen}
\email{ville.vaskonen@jyu.fi}
 \affiliation{University of Jyvaskyla, Department of Physics,\\
                      P.O.Box 35 (YFL), FI-40014 University of Jyvaskyla, Finland}
\affiliation{Helsinki Institute of Physics, \\
                      P.O.~Box 64, FI-00014, Helsinki, Finland}

\begin{abstract}
We consider an extension of the Standard Model with a singlet sector consisting of a real (pseudo)scalar and a Dirac fermion 
coupled with the Standard Model only via the scalar portal. 
We assume that the portal coupling is weak enough for the singlet sector not to thermalize with the Standard Model allowing the production of singlet particles via the freeze-in mechanism. 
If the singlet sector interacts with itself sufficiently strongly, 
it may thermalize within itself, resulting in dark matter abundance determined by the freeze-out mechanism operating within the singlet sector. We investigate this scenario in detail. In particular, we show that requiring the absence of inflationary isocurvature fluctuations provides lower bounds on the magnitude of the dark sector self-interactions and in parts of the parameter space favors sufficiently large self-couplings, supported also 
by the features observed in the small-scale structure formation.
\end{abstract}

\preprint{HIP-2016-9/TH}
%\pacs{pacs1,pacs2}

%%%%%%%%%%%%%%%%%%%%%%%%%%%%%%%%%%%%%%%%%%%%%%%%%%%%%%%%%%%%%%%%%%%%%%%%%%%%%%%%%%%%%%%%%%%%%%%%%%%%
% DOCUMENT
%
\maketitle
%%%%%%%%%%%%%%%%%%%%%%%%%%%%%%%%%%%%%%%%%%%%%%%%%%%%%%%%%%%%%%%%%%%%%%%%%%%%%%%%%%%%%%%%%%%%%%%%%%%%

%%%%%%%%%%%%%%%%%%%%%%%%%%%%%%%%%%%%%%%%%%%%%%%%%%%%%%%%%%%%%%%%%%%%%%%%%%%%%%%%%%%%%%%%%%%%%%%%%%%%
%%%%%%%%%%%%%%%%%%%%%%%%%%%%%%%%%%%%%%%%%%%%%%%%%%%%%%%%%%%%%%%%%%%%%%%%%%%%%%%%%%%%%%%%%%%%%%%%%%%%
%
\section{Introduction}

While the existence of a significant dark matter component in the universe has been clearly established, its interactions beyond the gravitational one remain to a large extent unknown. The direct search experiments provide stringent upper bounds for the coupling between dark matter and the Standard Model (SM) particles, while the astrophysical observations of galaxy cluster mergers provide limits for the possible self-interactions within the dark matter sector: for example, the Bullet Cluster requires $\sigma_{\rm DM}/m_{\rm DM} \lsim 1 \rm{cm}^2/\rm{g}$ \cite{Markevitch:2003at, Randall:2007ph, Rocha:2012jg, Peter:2012jh, Harvey:2015hha}.  If $\sigma_{\rm DM}/m_{\rm DM}$ is at the edge of this observational limit \cite{Kahlhoefer:2015vua, Campbell:2015fra}, one may still explain the shift of the gravitational centers observed for Abell 3827 \cite{Massey:2015dkw} and address the core-cusp problem \cite{Flores:1994gz,Navarro:1996gj,deBlok:2009sp}, the missing satellites problem \cite{Klypin:1999uc} and the too big to fail problem \cite{BoylanKolchin:2011de}.

In light of the direct search constraints, one appealing production mechanism for the dark matter abundance is the so-called 
freeze-in \cite{McDonald:1993ex, Hall:2009bx}. In this scenario, the coupling  between the dark and visible sectors takes a very small value, ${\cal O}(10^{-10})$, and the observed dark matter abundance is produced out-of-equilibrium from thermal bath of SM fields. 
However, strong dark matter self-interactions may lead to
a more complicated thermal history of the dark sector. It becomes necessary to consider the possibility of thermalization of the dark sector fields within themselves. This results in dark matter production due to the freeze-out mechanism which occurs in the dark sector. 

In this paper we consider, as a representative model example, the Higgs portal model with the dark sector constituted by a singlet scalar and a singlet fermion.
We investigate in detail the interplay between the constraints arising from the cosmological and astrophysical observations.
Earlier studies on observational properties of frozen-in dark matter in similar 
models include the case of an ultra-strongly interacting dark matter 
\cite{Pollack:2014rja}, cosmological, astrophysical and collider constraints on 
sterile neutrinos \cite{Roland:2015yoa,Merle:2015oja,Shakya:2015xnx, Merle:2015vzu}, 
and displaced signatures at colliders \cite{Co:2015pka, Ayazi:2015jij}. Frozen-in 
dark matter has also been used to explain the disagreement between structure 
formation in cold dark matter simulations and observations \cite{McDonald:2015ljz}. 
Dark matter interpretation of a spectral feature at $E\simeq 3.55$ keV observed in X-ray observations from several dark matter dominated sources \cite{Bulbul:2014sua,Boyarsky:2014jta} has been studied in \cite{Queiroz:2014yna,Baek:2014poa,
Farzan:2014foo,Arcadi:2014dca,Merle:2014xpa,Kolda:2014ppa}, and the galactic centre gamma ray 
excess in \cite{Heikinheimo:2014xza}.
In our analysis we will also comment on the parameter space relevant 
for the 3.55 keV line within our model.
Further cosmological constraints are discussed in e.g. \cite{Bernal:2015ova}.

Our main new result is that the constraint on isocurvature fluctuations, present in any dark sector not in thermal equilibrium with the visible sector
\cite{Markkanen:2015xuw, Kainulainen:2016vzv},  
favors non-negligible self-interactions in the dark sector. 
Combining this with the constraint from the 
astrophysical observations of cluster mergers then makes the possibility 
of nontrivial thermal history of the dark sector more plausible.

The paper is organized as follows: In Section \ref{abundance},
we introduce the model and 
consider the thermal history of the dark sector in detail.
In Section \ref{constraints}, we describe the scalar field dynamics in the early universe and 
present the analysis of the parameter space of the model 
imposing the cosmological and astrophysical constraints. In Section \ref{checkout}, 
we present our conclusions and outlook for further work. 

\section{The Model and the Relic Abundance}
\label{abundance}

As a concrete model example, we consider the Standard Model (SM) extended by a singlet sector consisting of a real singlet (pseudo)scalar $s$ and a sterile neutrino $\psi$. The only nongravitational interaction between this singlet sector and the visible SM sector is via the scalar portal coupling $\lambda_{\rm hs}|\Phi|^2s^2$, where $\Phi$ is the SM Higgs field.

We assume that the singlet sector is invariant under the parity transformation $\psi(t,x)\to\gamma^0\psi(t,-x)$ and $s(t,x) \to -s(t,-x)$. The fermionic part of the singlet sector Lagrangian is 
\be
\label{Lpsi}
\mathcal{L}_\psi = \bar\psi(i\slashed\partial - m_\psi)\psi + igs\bar{\psi}\gamma_5\psi~,
\ee
and the most general renormalizable scalar potential is given by
\be
V(\Phi,s) = \mu_{\rm h}^2\Phi^\dagger\Phi+\lambda_{\rm h}(\Phi^\dagger\Phi)^2+\frac{\mu_{\rm s}^2}{2} s^2+\frac{\lambda_{\rm s}}{4}s^4+\frac{\lambda_{\rm hs}}{2}\Phi^\dagger\Phi s^2 .
\label{potential}
\ee

In the vacuum at $T=0$, then, the mass of the singlet is $m_{\rm s}^2 = \mu_{\rm s}^2 + \lambda_{\rm hs}v_{\rm EW}^2/2$, where $v_{\rm EW}=246\mathrm{GeV}$ is the vacuum expectation value of the Higgs field. The mass of the Higgs particle is fixed to the observed value $m_{\rm h}=125$ GeV,
and we assume that $\mu_{\rm s}^2$ is positive and large enough so that $m_{\rm s}^2$ remains positive even if $\lambda_{\rm hs} < 0$.
We do not specify any particular dynamical origin for the scalar or fermion mass terms, as our goal is rather to investigate the observational consequences of this generic class of portal models.
Note that here either the fermion $\psi$ or the scalar $s$, or possibly both simultaneously, can be the dark matter candidate.

If the portal coupling takes a value $|\lambda_{\rm hs}| \lsim 10^{-7}$ the singlet sector does not enter thermal equilibrium with the visible sector \cite{Enqvist:2014zqa}
and the dark matter abundance is produced via the standard freeze-in scenario\footnote{Or alternatively, in the absence of the portal coupling, the dark sector could be populated via asymmetric reheating, as discussed in \cite{Adshead:2016xxj}.} which is our primary interest here. However, depending on the strength of the dark matter self-coupling, the thermal history of the dark sector may exhibit various features which we
shall now explore in more detail. 
Typically, when the freeze-in scenario is discussed, the effects of dark matter 
self-interactions are assumed to be negligibly small, in line with the feeble interactions connecting the dark matter to the SM. However, a nearly decoupled sector with nontrivial internal dynamics is a perfectly viable and technically natural scenario, as discussed in \cite{Foot:2013hna}. Moreover, we will show that the self-coupling of the frozen-in scalar dark matter field is bounded from below by the isocurvature constraints, further motivating the study of self-interacting frozen-in dark matter. The thermal history of a self-interacting dark matter species with number changing interactions, such as $3\rightarrow 2$ or $4\rightarrow 2$ scattering, was first discussed in \cite{Carlson:1992fn} and recently in e.g. \cite{Hochberg:2014dra, Pappadopulo:2016pkp}. The thermal history of a frozen-in self-interacting dark sector has been discussed in \cite{Chu:2011be, Bernal:2015ova} in the case where the dark sector is equipped with a massless force carrier, a dark photon. A scenario similar to ours has been studied in \cite{Bernal:2015xba}.

\subsection{Scalar freeze-out}

Let us consider first the scenario where the dark matter is constituted by the singlet scalar, $m_{\rm s}\le 2m_\psi$. For simplicity, we will assume that the fermion in this case is heavy and decoupled from the low-energy spectrum. The thermal history of the dark sector proceeds as follows: first, an initial population of dark matter is produced through Higgs decays
\cite{Chu:2011be},
\be
n_{\rm D}^{\rm initial} \simeq \left. 3\frac{n_{\rm h}^{\rm eq}\Gamma_{h\rightarrow ss}}{H}\right|_{T=m_{\rm h}},
\label{eq:n init}
\ee
where $n_{\rm h}^{\rm eq}$ is the equilibrium number density of the Higgs boson in the SM plasma, $H$ is the Hubble rate and the expression is evaluated when the temperature of the SM plasma is $T\approx m_{\rm h}$. The Higgs decay width into dark matter particles is given by
\be
\Gamma_{h\rightarrow ss} = \frac{\lambda_{\rm hs}^2 v_{\rm EW}^2}{32\pi m_{\rm h}}.
\label{eq:htoss}
\ee
In the standard freeze-in scenario this is the final relic abundance that simply dilutes with the scale factor after the production of dark matter through Higgs decays has stopped. However, if the number changing interactions, i.e. the $2\rightarrow 4$ scattering processes\footnote{The $2\rightarrow 3$ process is forbidden by the $\mathbf{Z}_2$-symmetry of the singlet scalar.} in the dark sector are fast, they will lead to a chemical equilibrium within the dark sector, reducing the average momentum of the dark matter particles and increasing their number density.

Threshold for thermalization of the dark sector is estimated by \mbox{$n_{\rm D} \langle \sigma_{2\rightarrow 4} v \rangle \gtrsim H $}. We further assume that the dark matter particles are much lighter than the Higgs, so that the initial dark matter population produced through Higgs decays is very relativistic. In the relativistic limit the $2\rightarrow 4$ scattering cross-section scales as \mbox{$\langle \sigma_{2\rightarrow 4} v \rangle  \sim T^{-2}\sim a^2$}, so that while the dark matter is relativistic, the scattering rate increases with respect to the Hubble rate as $n_{\rm D}\langle \sigma_{2\rightarrow 4} v \rangle/H\sim a$. As the SM bath temperature decreases below $T\sim m_{\rm s}$, the dark matter becomes nonrelativistic and the thermally averaged cross-section becomes a constant,
\be
\langle \sigma_{2\rightarrow 4} v \rangle \simeq \frac{\lambda_{\rm s}^4}{m_{\rm s}^2},
\ee
and the scattering rate starts to decrease with respect to the Hubble rate as $\sim a^{-1}$. Therefore, threshold for thermalization of the dark sector can be estimated by comparing the scattering rate to the Hubble rate at temperature $T=m_{\rm s}$. We find that, within this approximation, the thermalization will take place if the self-coupling exceeds the critical value
\be
\lambda_{\rm s}^{\rm FI} \simeq \sqrt{\frac{52.7(g_*(m_{\rm h})g_*(m_{\rm s}))^\frac14 \sqrt{m_{\rm h} m_{\rm s}}}{\lambda_{\rm hs}M_{\rm P} }},
\ee
where $g_*(T)$ is the effective number of relativistic degrees of freedom in the SM plasma at temperature $T$ and $M_{\rm P}$ is the Planck mass. For $\lambda_{\rm{s}}<\lambda_{\rm{s}}^{\rm{FI}}$ the usual freeze-in picture is sufficient.

If the self-coupling is larger than this value, the dark sector enters
chemical equilibrium, where the $2\leftrightarrow 4$ interactions 
maintain the equilibrium number density 
at the dark sector temperature $T_{\rm D}$, until the $4\rightarrow 2$ interaction rate drops below the Hubble rate and the number density freezes out. This mechanism is referred to as {\it dark freeze-out}. The final relic abundance
depends on the freeze-out temperature of the $4\rightarrow 2$ scattering rate, which
in the nonrelativistic limit is estimated as
\be
\langle \sigma_{4\rightarrow 2} v\rangle \simeq \frac{\lambda_{\rm s}^4}{m_{\rm s}^8},
\ee
and on ratio of temperatures in the dark and visible sectors.
This is given by the initial energy density transferred to the dark sector through the Higgs decays as
\be
\xi = \frac{T_{{\rm D}}}{T} = \left( \frac{\rho_{\rm D}}{ g_{*{\rm D}}}\frac{g_* }{\rho}\right)^\frac14,
\label{eq:chidef}
\ee
where $g_{*D}=2$ is the number of relativistic degrees of freedom of the dark sector, $\rho$ is the energy density of the SM plasma, \mbox{$\rho_{\rm D} \simeq \Gamma_{h\rightarrow ss}(T)m_{\rm s}n_{\rm h}^{\rm eq}(T)/H(T)$} is the energy density of the dark sector, where \mbox{$\Gamma_{h\rightarrow ss}(T) = \Gamma_{h\rightarrow ss}K_1(m_{\rm h}/T)/K_2(m_{\rm h}/T)$} \cite{Chu:2011be}.

After the decoupling of the visible and dark sectors at $T\sim m_{\rm h}/3$, the entropies of both sectors are conserved separately, and hence the ratio of the entropy densities $\chi\equiv s/s_{\rm D}$ remains constant. As was first derived in \cite{Carlson:1992fn}, this can be cast as an equation for the dark freeze-out temperature $x_{\rm D}^{\rm FO} = m_{\rm s}/T_{\rm D}^{\rm FO}$ in terms of the current dark matter abundance $\Omega_{\rm DM}$:
\be
x_{\rm D}^{\rm FO} = \frac{m_{\rm s}}{3.6\ {\rm eV}\ \Omega_{\rm DM} h^2 \chi}.
\label{eq:xD1}
\ee
By entropy conservation, $\chi$
can be expressed in terms of the initial value $\xi_0$ 
obtained by evaluating Eq. (\ref{eq:chidef}) at $T_0= m_{\rm h}/3$.

On the other hand, the dark freeze-out temperature can be estimated as the temperature at which the $4\rightarrow 2$ interaction rate drops below the Hubble rate, resulting in
\be
x_{\rm D}^{\rm FO} = \frac13 \log\left(\left(\frac{1}{2\pi}\right)^\frac92 \frac{\xi^2\lambda_{\rm s}^4 M_{\rm P}}{1.66\sqrt{g_*}m_{\rm s} (x_{\rm D}^{\rm FO})^\frac52 }\right), 
\label{eq:xD2}
\ee

\begin{figure}
\begin{center}
\includegraphics[width = 0.47\textwidth]{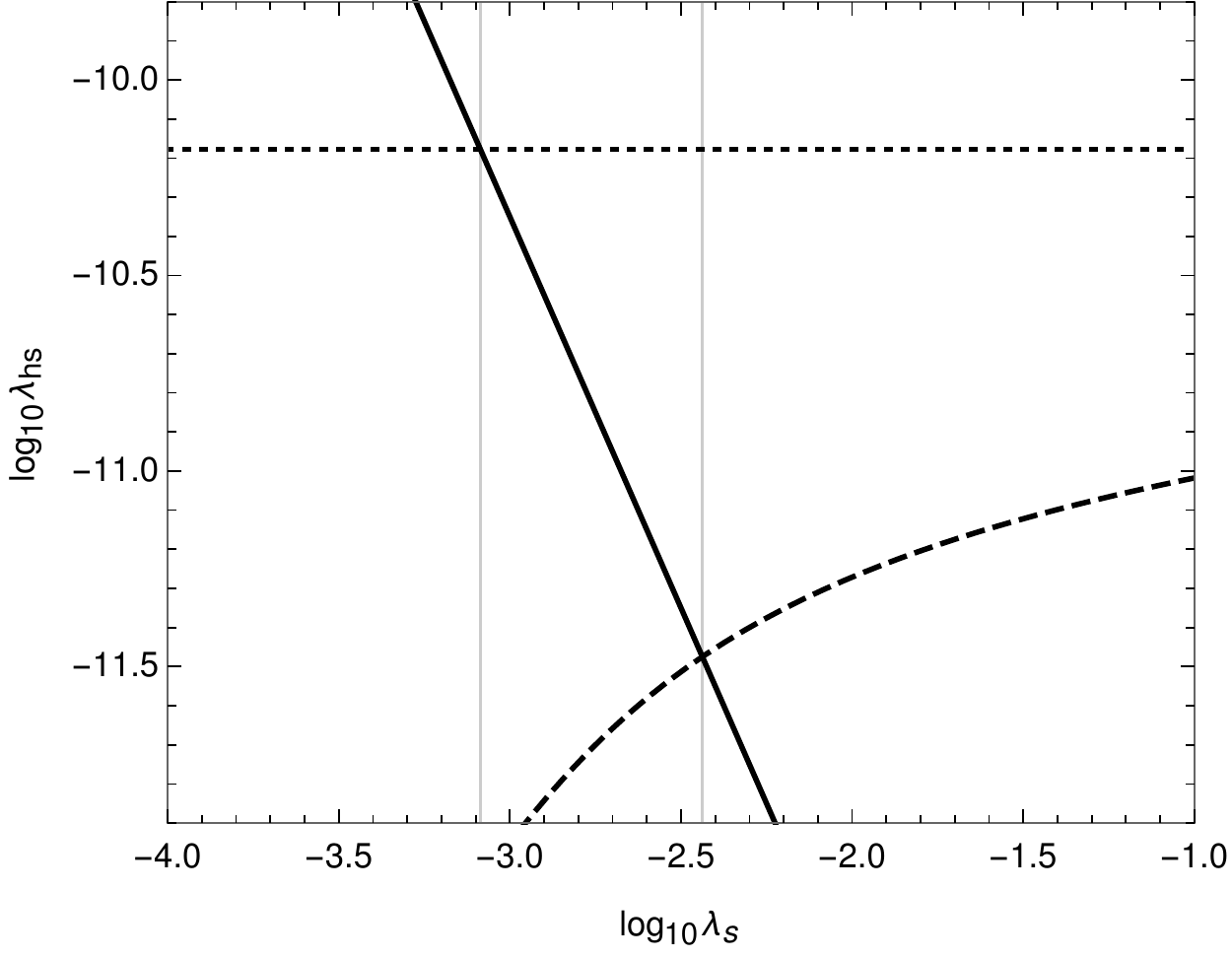}
\caption{The portal coupling that yields the correct dark matter abundance as a function of the dark matter self-coupling for $m_{\rm s}=0.1$ GeV, in the thermal i.e. dark freeze-out scenario (dashed line) and in the nonthermal i.e. in the standard freeze-in scenario (dotted horizontal line). The solid diagonal line is the critical value for thermalization of the dark sector.}
\label{fig: thermal vs nonthermal}
\end{center}
\end{figure}

Equating (\ref{eq:xD1}) with (\ref{eq:xD2}) and requiring the dark matter abundance to be the observed one, $\Omega_{\rm DM}h^2=0.12$, yields a relation between the three parameters of the model, $m_{\rm s}, \lambda_{\rm s}, \lambda_{\rm hs}$. An example of how the dynamics we have discussed here operates is depicted in Figure \ref{fig: thermal vs nonthermal} for fixed value of the dark matter mass, $m_{\rm s}=0.1$ GeV. The dashed line in the lower right corner shows the portal coupling that results in the observed dark matter abundance in the dark freeze-out scenario. The dotted horizontal line is the value that produces the correct relic abundance in the standard freeze-in scenario, where the dark sector does not thermalize. The critical value for thermalization is presented by the solid diagonal line, above which the dark sector reaches chemical equilibrium. 
To the left of the solid line the relic abundance can be obtained by the standard freeze-in mechanism, and to the right by the dark freeze-out mechanism. 
In the region between the two vertical gray lines, there are no solutions that would yield the correct relic abundance within the approximations made in this calculation.

\begin{figure}
\begin{center}
\includegraphics[width = 0.45\textwidth]{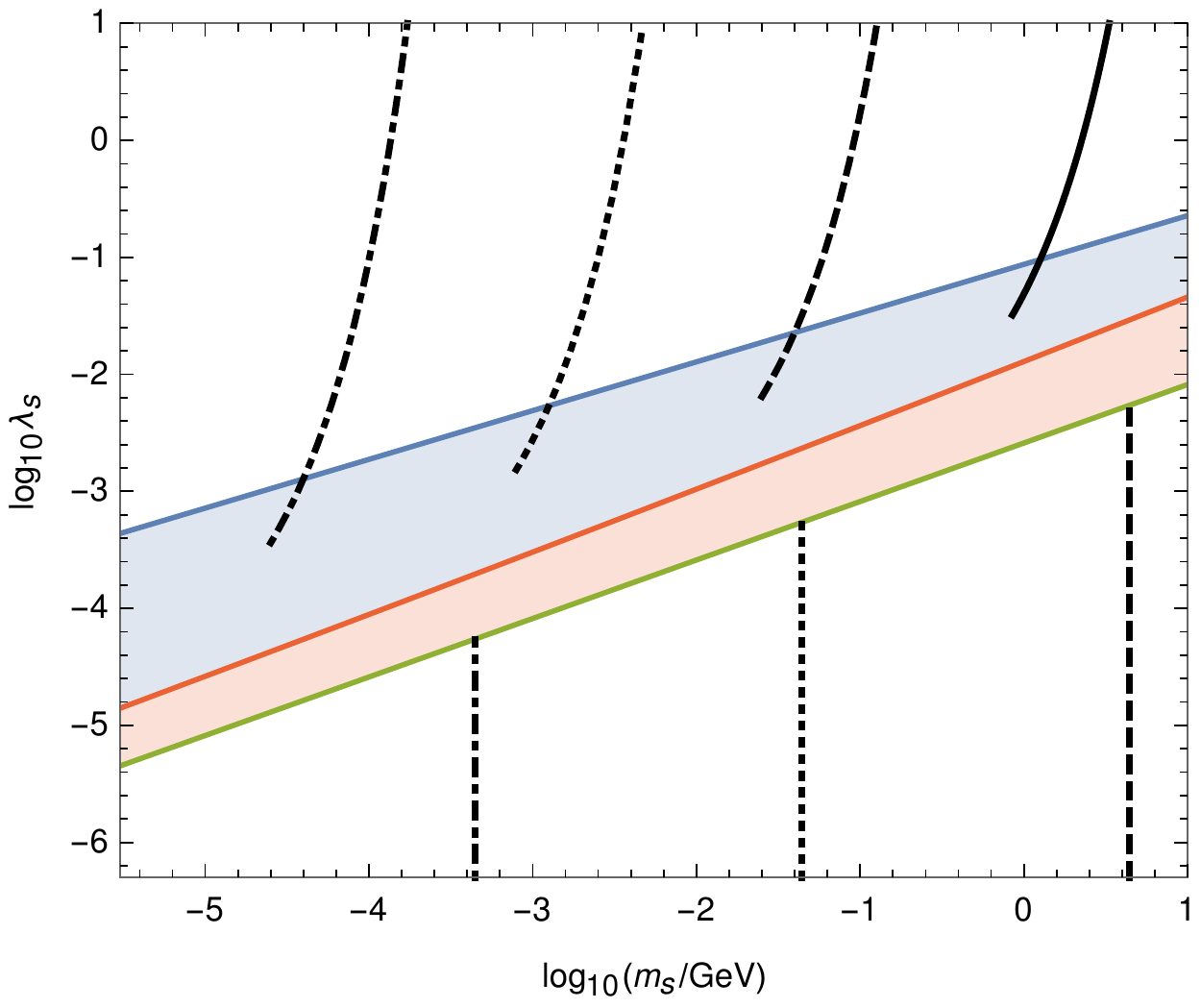}
\caption{The self-coupling as a function of the dark matter mass required to obtain the correct relic abundance for $\lambda_{\rm hs}=10^{-12}$, (solid black line) $\lambda_{\rm hs}=10^{-11}$ (dashed black line), $\lambda_{\rm hs}=10^{-10}$ (dotted black line) and $\lambda_{\rm hs}=10^{-9}$ (dot-dashed black line). In the blue shaded region the freeze-out happens at a (semi)relativistic temperature, 
in the red shaded region we find no solutions to obtain the observed dark matter abundance, and in the lower white region the dark matter abundance can be obtained via the standard freeze-in scenario. 
}
\label{fig: freeze in yield}
\end{center}
\end{figure}

The results as a function of $m_{\rm s}$ and $\lambda_{\rm s}$ are shown in Figure 
\ref{fig: freeze in yield}.
Along the solid, dashed, dotted and dot-dashed black lines the correct relic density is obtained for portal couplings $\lambda_{\rm hs}=10^{-12}$, $10^{-11}$, $10^{-10}$ and $10^{-9}$, respectively.
Above the blue shaded region the abundance is produced by the dark freeze-out mechanism at a temperature where the dark matter is non-relativistic. Within the blue shaded region 
the dark freeze-out happens at a (semi)relativistic temperature,
characterized by $x_{\rm D}^{\rm FO} \leq 3$. In the red region below the blue one, we find no solutions that would yield the correct relic abundance; see Figure \ref{fig: thermal vs nonthermal}.
Below the red region the observed dark matter abundance can be produced via the usual freeze-in mechanism.

The allowed parameter space for the dark matter mass is limited from below by the constraints on warm dark matter arising from Lyman-$\alpha$ forest data \cite{Viel:2013apy}, excluding warm dark matter with mass below $m_{\rm DM}\approx 3\ {\rm keV}$. Relativistic dark freeze-out for dark matter mass above this limit, however, is compatible with constraints from structure formation and matter power spectrum. In Figures \ref{fig: freeze in yield} and \ref{fig: freeze in yield fermion} the blue region therefore represents allowed parameter space. However, our calculation for the relic abundance has been performed in the limit of nonrelativistic freeze-out, and therefore the required value of the portal coupling for obtaining the correct dark matter abundance is subject to relativistic corrections, which are not generally small in this region. Therefore, we have truncated the lines presenting the solution for the selected values of $\lambda_{\rm hs}$ in this region. A solution that yields the correct dark matter abundance can in general be found also in the relativistic case, but we postpone the detailed numerical analysis of this problem for later work. Furthermore, it should be noted that the red region where our method of approximating the dark matter abundance produces no valid solutions, lies deep in the relativistic freeze-out regime. Therefore it is likely that also within this region, a full numerical analysis would find a solution that produces the observed relic abundance.

\subsection{Fermionic freeze-out}

Including the fermion in the spectrum, we can essentially repeat the previous treatment for the relic abundance. We will focus here on the scenario where a mass hierarchy $m_\psi\ll m_{\rm s} \ll m_{\rm h}$ applies, so that the initial production of dark matter proceeds through the Higgs decays into the scalars $s$ as above, followed by the decay $s\rightarrow \bar{\psi}\psi$. 

Again, the thermal evolution of the dark sector is very sensitive to the self-interactions in the scalar sector. Due to the presence of the fermion Yukawa coupling the situation is also somewhat different from the scalar case treated in the previous section. Depending on the relative strength of the scalar $2\rightarrow 4$ scattering rate and the decay rate
\be
\Gamma_{{\rm s}\rightarrow\bar{\psi}\psi} = \frac{g^2 m_{\rm s}}{8\pi},
\ee
the scalars may thermalize before their decay. 

The ratio of the scattering and decay rates can be estimated as
\be
\frac{\gamma^{-1} \Gamma_{{\rm s}\rightarrow\bar{\psi}\psi}}{n_{\rm D}^{\rm initial}\left. \langle\sigma_{2\rightarrow 4}v\rangle \right|_{T\sim \frac12 m_{\rm h}}} \simeq 1.1\sqrt{g_*(m_{\rm h})} \left(\frac{g^2}{\lambda_{\rm hs}^2\lambda_{\rm s}^4}\right)
\left(\frac{m_{\rm h}}{M_{\rm P}}\right),
\label{eq: dec/scatter}
\ee
where $\gamma = m_{\rm h}/2m_{\rm s}$ is the relativistic time dilation factor.

Consider first the case where the above ratio is large, i.e. the part of the parameter space where the scalar scattering is irrelevant for the thermalization of the dark sector. The relevant scattering processes are  
$2\rightarrow 4$ fermion scatterings, and in the nonrelativistic limit their cross-sections are
\be
\begin{split}
\langle \sigma_{2\rightarrow 4} v \rangle &\simeq \frac{g^8}{m_{\rm s}^2},  \\
\langle \sigma_{4\rightarrow 2} v \rangle &\simeq \frac{g^8}{m_{\rm s}^8}\frac{m_\psi^2}{m_{\rm s}^2}.
\end{split}
\ee
We can then repeat the analysis of the previous section, keeping in mind that the number of relativistic degrees of freedom contributing to the initial entropy density of the dark sector is now $g_{*D}=2+\frac74$ for one charged scalar and one Dirac fermion. We find that the critical coupling above which the dark sector thermalizes is now given as
\be
g^{\rm FI} \simeq \left( \frac{37.2(g_*(m_{\rm h})g_*(m_\psi))^\frac14 \sqrt{m_{\rm h}} m_{\rm s}}{\lambda_{\rm hs}M_{\rm P}\sqrt{m_\psi}} \right)^\frac14,
\ee
and the freeze-out temperature of the $4\rightarrow 2$ scattering as
\be
x_{\rm D}^{\rm FO} = \frac13 \log\left( \left(\frac{1}{2\pi}\right)^\frac92 \frac{\xi^2 g^8 M_{\rm P} m_\psi^9}{1.66\sqrt{g_*}m_{\rm s}^{10}(x_{\rm D}^{\rm FO})^\frac52 }\right),
\ee
where $x_{\rm D}^{\rm FO} = m_{\rm \psi}/T_{\rm D}^{\rm FO}$.

\begin{figure}
\begin{center}
\includegraphics[width = 0.45\textwidth]{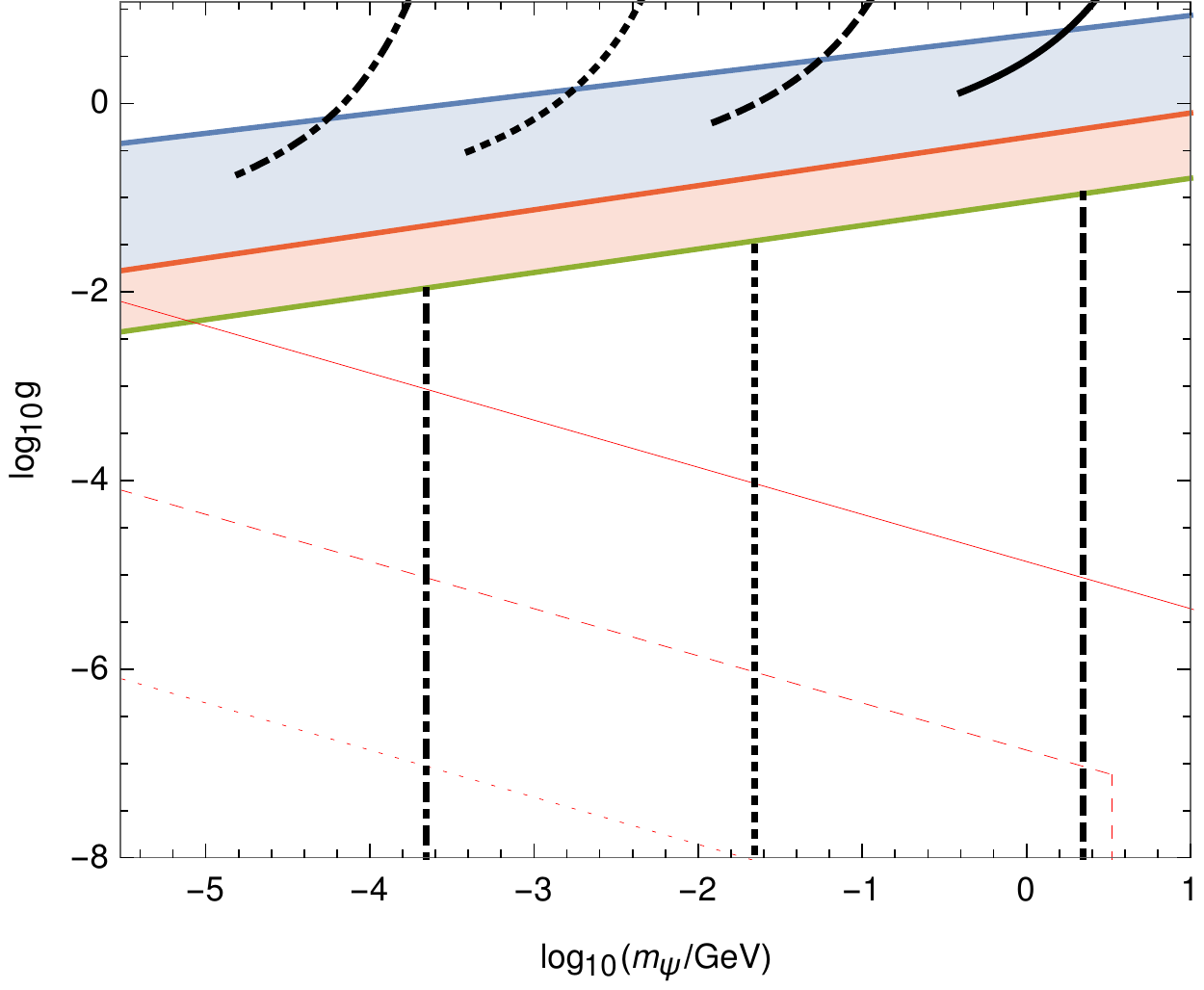}
\caption{The Yukawa coupling as a function of the fermion mass required to obtain the correct relic abundance for \mbox{$\lambda_{\rm hs}=10^{-12}$}, (solid black line) $\lambda_{\rm hs}=10^{-11}$ (dashed black line), $\lambda_{\rm hs}=10^{-10}$ (dotted black line) and $\lambda_{\rm hs}=10^{-9}$ (dot-dashed black line) assuming $m_{\rm s}=10m_{\psi}$. The color coding of the shaded regions is the same as in Figure \ref{fig: freeze in yield}. Below the solid, dashed and dotted red contours the dark sector thermalizes via scalar self-scattering before the scalars have decayed to fermions, for $\lambda_{\rm s} = 0.1, 0.01$ and $0.001$, respectively.}
\label{fig: freeze in yield fermion}
\end{center}
\end{figure}

Since the number changing fermion self-scattering is suppressed by 8 powers of the Yukawa coupling, the fermionic dark sector thermalizes only for relatively large values of $g$. This is shown in Figure \ref{fig: freeze in yield fermion}, where the color coding of the thick black lines and the shaded regions is as in Figure \ref{fig: freeze in yield}, and we have fixed the mass hierarchy of the dark sector as $m_{\rm s} = 10 m_\psi$. From Figure \ref{fig: freeze in yield fermion} we see that for $g\lesssim 0.01$ the dark sector does not thermalize via the above mechanism, and the relic abundance may be achieved by the usual freeze-in mechanism. 

However, let us now take into account that if the scalar self-coupling is large enough, i.e. the ratio in Eq. (\ref{eq: dec/scatter}) is small, the dark sector can thermalize immediately after the initial production of the scalars by scalar $2\rightarrow 4$ scattering.
This boundary is depicted in Figure \ref{fig: freeze in yield fermion} by the solid, dashed and dotted red contours corresponding, respectively, to $\lambda_{\rm s} = 0.1, 0.01$ and $0.001$.
In the region below and to the left of these contours 
the dark sector will thermalize and affect the determination of the relic abundance.

\begin{figure}
\begin{center}
\includegraphics[width = 0.45\textwidth]{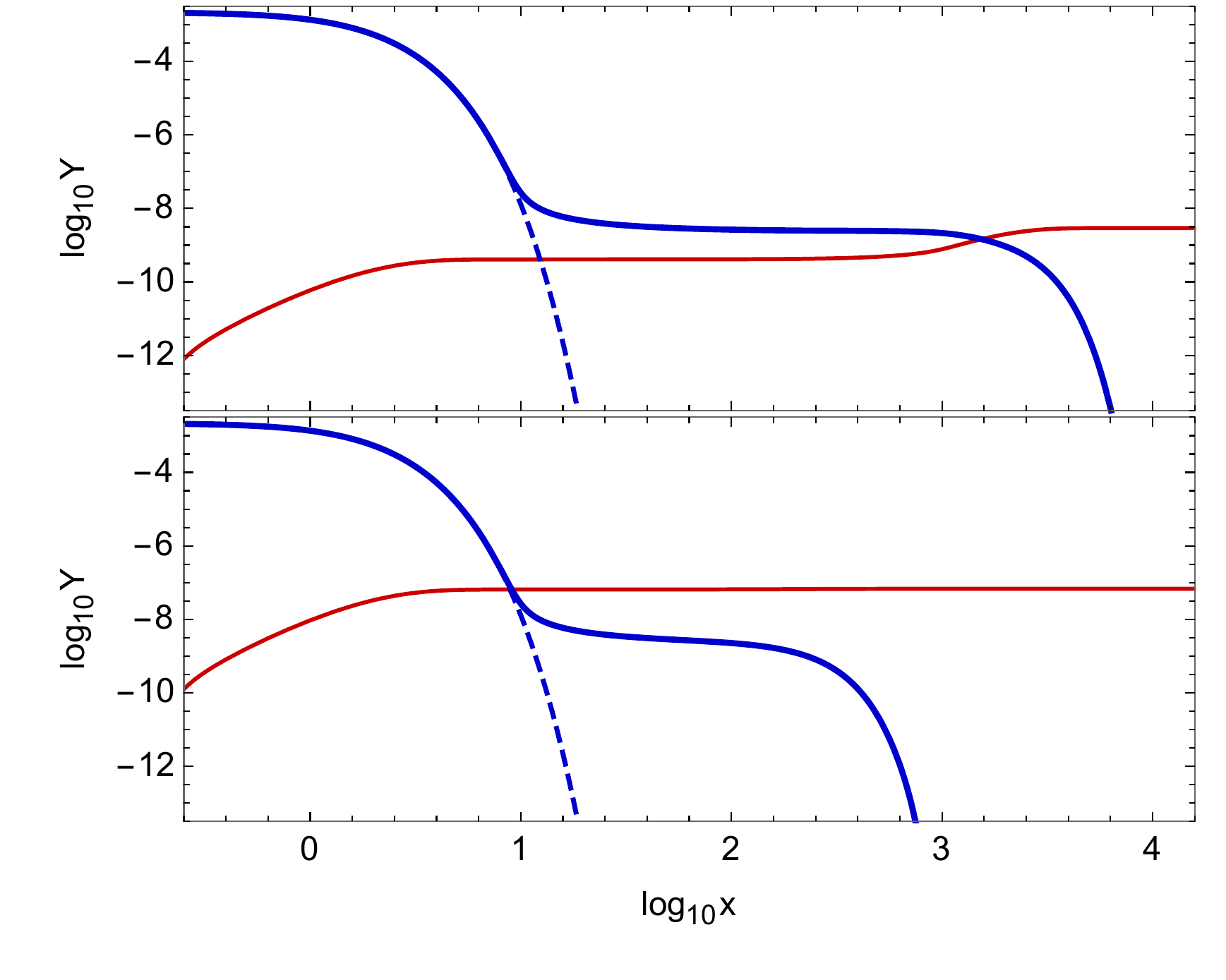}
\caption{Upper panel: if the fermion Yukawa coupling is very small, the particle production is dominated by the dark freeze-out of the scalar abundance (blue thick curve). As $2m_\psi\le m_{\rm s}$ these will ultimately decay to fermions and the dark freeze-in will provide only a subleading contribution to the fermion abundance (red thin curve). Lower panel: as the Yukawa coupling increases, but is not strong enough for the fermions to thermalize, the dark freeze-in production (red thin curve) is more important than the contribution from the scalar freeze-out (blue thick curve).}
\label{fig:boltzmann}
\end{center}
\end{figure}

In this region, there are different possibilities depending on how thermalization is attained between different particle species. Two different possibilities for particle production dynamics are qualitatively depicted in 
Figure \ref{fig:boltzmann}: first, the upper panel of the Figure corresponds to the case where the fermion Yukawa coupling $g$ is very small, and the relic abundance is determined by the scalar freeze-out. The scalars have a long lifetime, $\tau\sim 1/g^2$, and ultimately they decay to fermions. Due to smallness of the coupling $g$, particle production via fermion freeze-in during the phase where scalars have thermalized is negligible. Second, the lower panel of the Figure corresponds to the case where the coupling $g$ is small enough for the fermions not to thermalize but large enough for the dark freeze-in of fermions to become the dominant dark matter production mechanism. In this case most of the final dark matter abundance is produced out-of-equilibrium within the dark sector, from the thermal bath of the singlet scalars, in an analogous manner to the usual freeze-in mechanism. For parameter values corresponding to the domain directly below the red lines of Figure \ref{fig: freeze in yield fermion}, thermalization of fermions becomes also possible. We leave the detailed investigation of these dynamics, in particular the region
where the ratio (\ref{eq: dec/scatter}) is of order one, for future work, and consider in this paper the region 
above the red lines in Figure \ref{fig: freeze in yield fermion}, where the ratio is large.

\section{Observational constraints}
\label{constraints} 

\subsection{Scalar field dynamics in the early universe}

In order to fully
utilize the cosmological and astrophysical 
constraints, it is crucial to study particle dynamics not only in the vicinity of the current collider energies but at the highest energy scales we can probe also by other means, i.e. the scales of cosmic inflation. 

Scalar fields which are light during inflation {\it typically} acquire large fluctuations proportional to the inflationary scale $H$. The resulting displacement of the field from its vacuum creates an effective primordial condensate over the observable universe \cite{Starobinsky:1994bd}. Such primordial condensate of inflationary origin in general constitutes an isocurvature component uncorrelated with perturbations in the SM sector. For scalars very weakly coupled with the SM, this isocurvature component will persist. Furthermore, if such scalar sources the dominant dark matter component, the isocurvature fluctuations get imprinted on the dark matter abundance. Such a scenario is strictly constrained by observations of the Cosmic Microwave Background (CMB). 

In our model, the singlet scalar $s$ and the Higgs field acquire large inflationary fluctuations. The Higgs condensate decays rapidly into SM particles, see \cite{Enqvist:2014zqa}, and any isocurvature component is washed away. On the other hand, the singlet field does not feel the thermal bath of SM particles if $|\lambda_{\rm hs}| \lsim 10^{-7}$ and, consequently, it can produce a sizeable fraction of dark matter out of the primordial $s$ condensate \cite{Nurmi:2015ema, Kainulainen:2016vzv}. 

Let us consider the generation and evolution of the fluctuation $\delta s\sim H$ in more detail. Assuming $|\lambda_{\rm hs}|\ll \sqrt{\lambda_{\rm h}\lambda_{\rm s}}$, the probability of having a particular initial value $s_*$ at the onset of the post-inflationary era over the entire observable universe is described by the probability distribution function \cite{Starobinsky:1994bd}
\be
\label{distribution}
P_{\rm s}(s)=N {\rm exp}\left(-\frac{8\pi^2 V(s)}{3H^4}\right),
\ee
where $N$ is a normalization constant and \mbox{$V(s)=\mu_{\rm s}^2/2\,s^2+\lambda_{\rm s}/4\,s^4$}. A typical value for $s_*$ is
\be
\label{typical_s}
\sqrt{\langle s^2 \rangle}= 
\begin{cases}
0.363 H_*/\lambda_{\rm s}^{\scriptscriptstyle 1/4} \\
0.195 H_*^2/\mu_{\rm s} ,
\end{cases} 
\ee
depending, respectively, on whether the quartic or quadratic term dominates the potential at the end of inflation. Here $H_*$ denotes the value of the Hubble parameter at the horizon crossing of largest observable modes.

If $\lambda_{\rm s} s_*^2 > \mu_{\rm s}^2$, the quartic term dominates the potential at the end of inflation. Using the typical values \eqref{typical_s} for $s_*$ we see that 
this condition requires $\lambda_{\rm s}\gsim 10 (\mu_{\rm s}/H_*)^4$, i.e. for the quadratic term to dominate the potential at the end of inflation either $\lambda_{\rm s}$ has to be very small or $\mu_{\rm s}\sim H_*$. Because in astrophysical considerations the mass of the dark matter particle is usually required to be very small compared to the inflationary scale, in the following we will consider only the case where the quartic term dominates the potential at the end of inflation.

After inflation, the $s$ field remains nearly constant until it becomes effectively massive, $V'' \sim H^2$. After this, the homogeneous $s$ condensate starts to oscillate with a decreasing envelope.
The primordial $s$ condensate can then decay to dark matter particles which never come into thermal equilibrium with SM particles if $|\lambda_{\rm hs}|\lesssim 10^{-7}$ \cite{Nurmi:2015ema,Kainulainen:2016vzv}. The dark matter abundance sourced by a primordial field is given by 
\be
\label{highT_abundance}
\frac{\Omega_{\rm DM}^{({\rm s}_0)} h^2}{0.12} \simeq 
3.4\times10^{-4} n \lambda_{\rm s}^{\scriptscriptstyle -1/4} \left(\frac{m_{\rm DM}}{\mathrm{GeV}}\right) \left(\frac{s_*}{10^{11}\mathrm{GeV}}\right)^{\scriptscriptstyle 3/2} ,
\ee
where $n=1$ if the primordial field decays to $s$ particles or if the primordial field does not decay before photon decoupling, and $n=2$ if the primordial field decays to fermions directly or via process $s_0\to 2s\to 4\psi$. Whether the primordial field decays to scalars or to fermions depends on the couplings, see reference \cite{Kainulainen:2016vzv} for details.

\subsection{Isocurvature from a primordial source}

The dark matter component sourced by the primordial field constitutes isocurvature fluctuations. The amplitude of isocurvature fluctuations is heavily constrained by CMB observations, which give an absolute upper bound for dark matter abundance sourced by the singlet condensate \cite{Kainulainen:2016vzv}
\be
\label{icbound}
\frac{\Omega_{\rm DM}^{({\rm s}_0)} h^2}{0.12} \lsim 4.5\times10^{-5}\frac{s_*}{H_*}.
\ee
The singlet particles can still constitute all dark matter when most of the abundance is produced by the standard freeze-in mechanism and only a small fraction by decay of the primordial field.

To see how probable it is for the given values of model parameters ($H_*$, 
$m_{\psi}$, $\lambda_{\rm s}$ and $g$) to satisfy the isocurvature constraint \eqref{icbound}, we calculate the probability distribution function for $f(s_*)\equiv \Omega_{\rm DM}^{({\rm s}_0)} h^2/0.12/(4.5\times10^{-5} s_*/H_*)$,
\be
P_{\rm f}(z) = 2\left|\frac{\rm d}{{\rm d} z} f^{-1}(z)\right| P_{\rm s}(f^{-1}(z)),
\ee
where $P_{\rm s}$ is given by \eqref{distribution}. Now the integral \mbox{$P\equiv \int_0^1{\rm d}z P_{\rm f}(z)$} gives the probability of having a small enough initial field value $s_*$ to get below the isocurvature constraint \eqref{icbound}.

Using \eqref{highT_abundance}, 
and requiring $P>0.1\%$, i.e. that we do not live in a very atypical universe, we get an upper bound on dark matter particle mass
\be
\label{isocurvature}
\frac{m_{\rm DM}}{\rm GeV} \lsim \frac{6}{n}\lambda_{\rm s}^{\scriptscriptstyle 3/8}\left(\frac{H_{*}}{10^{11}{\rm GeV}}\right)^{\scriptscriptstyle -3/2} ,
\ee
which could also be expressed as a lower bound on scalar self-interaction strength for fixed $m_{\rm DM}, H_*$.

\subsection{Dark matter self-interactions}

As shown in Eq. (\ref{isocurvature}), cosmological constraints imply a lower bound on scalar self-interaction strength. On the other hand, astrophysical observations provide an upper bound on dark matter self-interactions. Assuming that all dark matter is self-interacting, the upper bound is \cite{Markevitch:2003at, Randall:2007ph, Rocha:2012jg, Peter:2012jh, Harvey:2015hha}
\be
\label{selfint_bound}
\frac{\sigma_{\rm DM}}{m_{\rm DM}}\lsim 1{\rm \frac{cm^2}{g}}.
\ee 
In the limit $m_{\rm s}\ll m_{\rm h}$ the singlet scalar self-interaction cross-section divided by its mass is
\be
\label{scrosssection}
\frac{\sigma_{\rm s}}{m_{\rm s}} = \frac{9\lambda_{\rm s}^2}{32 \pi m_{\rm s}^3}.
\ee 

If $s$ constitutes dominant fraction of total dark matter abundance, we get a lower bound on $s$ mass by combining \eqref{selfint_bound} and \eqref{scrosssection}. Hence the dark matter particle mass is bounded from above by the isocurvature bound \eqref{isocurvature} and from below by \eqref{selfint_bound},
\be
\label{ms_ubound}
0.026\lambda_{\rm s}^{\scriptscriptstyle 2/3} \lsim \frac{m_{\rm s}}{\rm GeV} \lsim 6\lambda_{\rm s}^{\scriptscriptstyle 3/8}\left(\frac{H_{*}}{10^{11}{\rm GeV}}\right)^{-3/2}.
\ee

\begin{figure}
\begin{center}
\includegraphics[width=.45\textwidth]{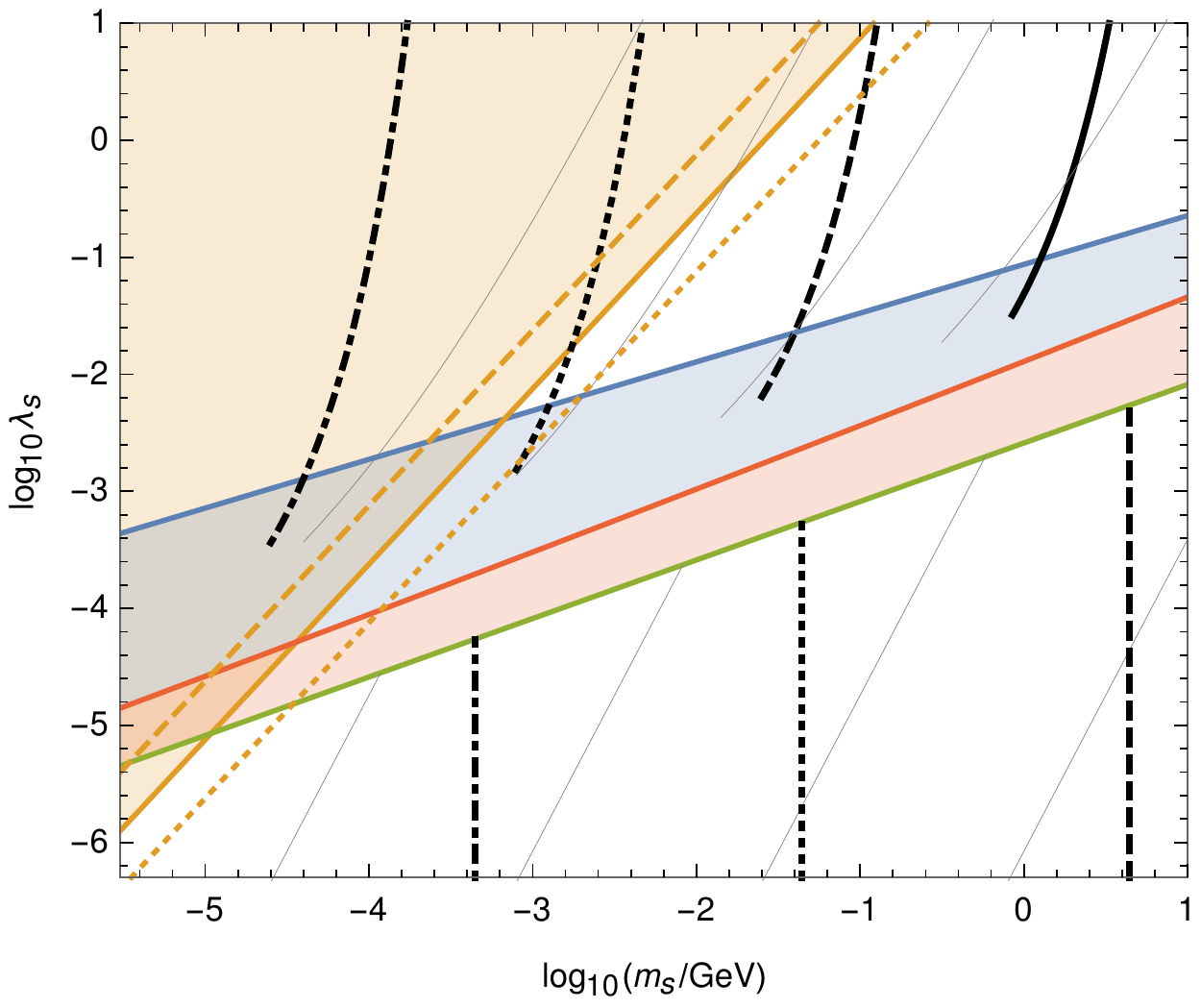}
\caption{The self-interaction bound and isocurvature constraints for the scalar dark matter scenario. The self-interaction limit, $\sigma_{\rm DM}/m_{\rm DM} < 1$ cm$^2$/g, is shown by the yellow shaded region in the top left corner together with the $\sigma_{\rm DM}/m_{\rm DM} = 10, 0.1$ cm$^2$/g contours (dashed and dotted, respectively), and the isocurvature constraints by the gray contours for \mbox{$H_* = 10^{13}, 10^{12}, 10^{11}, 10^{10}$ GeV} from left to right. The area to the right of each contour is ruled out for the given value of the inflationary scale $H_*$. The blue and red shaded regions and the black lines correspond to those in Figure \ref{fig: freeze in yield}.}
\label{fig2}
\end{center}
\end{figure}

These bounds are illustrated in Figure ~\ref{fig2}, superimposed over the relic abundance limits discussed in Figure \ref{fig: freeze in yield}. The yellow shaded region in the top left corner is constrained by the self-interaction bound \mbox{$\sigma_{\rm DM}/m_{\rm DM} < 1$ cm$^2$/g} from the cluster merger observations, while the dotted and dashed yellow lines show the astrophysically interesting region corresponding to the self-interaction cross-section between 0.1 and 10 cm$^2$/g. The gray contours show the isocurvature constraint for \mbox{$H_* = 10^{13}, 10^{12}, 10^{11}, 10^{10}$ GeV} from left to right, so that for a given inflationary scale the region to the right of these contours is ruled out by the non-observation of isocurvature fluctuations. For example, for Higgs inflation $H_*\simeq 2\times 10^{13}$ GeV \cite{Bezrukov:2007ep}, which, if verified by the next generation CMB satellites, rules out large portions of the parameter space.

The isocurvature constraint (\ref{isocurvature}) 
was derived in \cite{Kainulainen:2016vzv} assuming 
that the comoving number densities of the singlet scalars produced by 
the decay of the primordial condensate and via the freeze-in mechanism 
are separately conserved. It is then straightforward to compare the abundance of the primordial component that contributes to isocurvature fluctuations to the overall abundance of dark matter.
However, if the number changing interactions in the dark sector become 
active and the dark sector thermalizes, the situation is slightly more complicated. 
In this case we have derived the bound by comparing the energy density carried by the
scalars produced from the primordial condensate to the energy density of the scalars 
produced via the freeze-in mechanism, at the time of the dark sector thermalization. For simplicity, we assume thermalisation to take place at $T=m_{\rm s}$, which is the 
latest moment when the dark sector can reach chemical equilibrium, as discussed above. Using $\rho\sim a^{-3}$ we evaluate the energy density of the component carrying the isocurvature contribution, equation \eqref{highT_abundance}, at the thermalisation temperature. The adiabatic component, produced via the freeze-in mechanism, is relativistic before thermalization, so the initial energy density given by \eqref{eq:n init} scales as $\rho\sim a^{-4}$ from the initial freeze-in temperature $T\approx \frac12 m_h$ to the thermalization temperature $T\approx m_{\rm s}$.

After the equilibration of the dark sector, the particles from both origins 
will contribute to the thermal bath of dark matter, so that the relative abundance of 
the isocurvature component with respect to the total dark matter abundance will 
remain constant from there on and we can directly apply the result \eqref{icbound} to evaluate the isocurvature constraint in this case. 

The effect of this correction is to increase the importance of the isocurvature constraint: thermalization of the dark sector increases the number density of dark matter particles, resulting in a larger final dark matter abundance than in the standard freeze-in scenario, so that in order to produce the observed dark matter abundance in the end, a smaller initial abundance of scalars is needed. Thus, an initial population of scalars produced from the decay of the primordial condensate will contribute a larger fraction of the total dark matter energy density than it would in the standard freeze-in scenario. 

As our calculation of the initial dark matter abundance required in the dark freeze-out scenario has been performed assuming a non-relativistic freeze-out, we have truncated also the isocurvature contours in the blue shaded region, where our solution would be subject to large relativistic corrections. We shall postpone a more detailed numerical analysis of this situation for a future publication.

Overall, the noticable feature is that, in the case of scalar dark matter of mass $m_{\rm s}$ and for fixed inflationary scale $H_*$, the isocurvature constraint bounds the scalar self-interactions from below, possibly resulting in nontrivial thermal history of the dark sector as discussed in the previous section. For a very large inflationary scale $H_*$ both the mass $m_{\rm s}$ and the self-interaction coupling $\lambda_{\rm s}$ have to be very small to satisfy the bounds.

\begin{figure}
\begin{center}
\includegraphics[width=.45\textwidth]{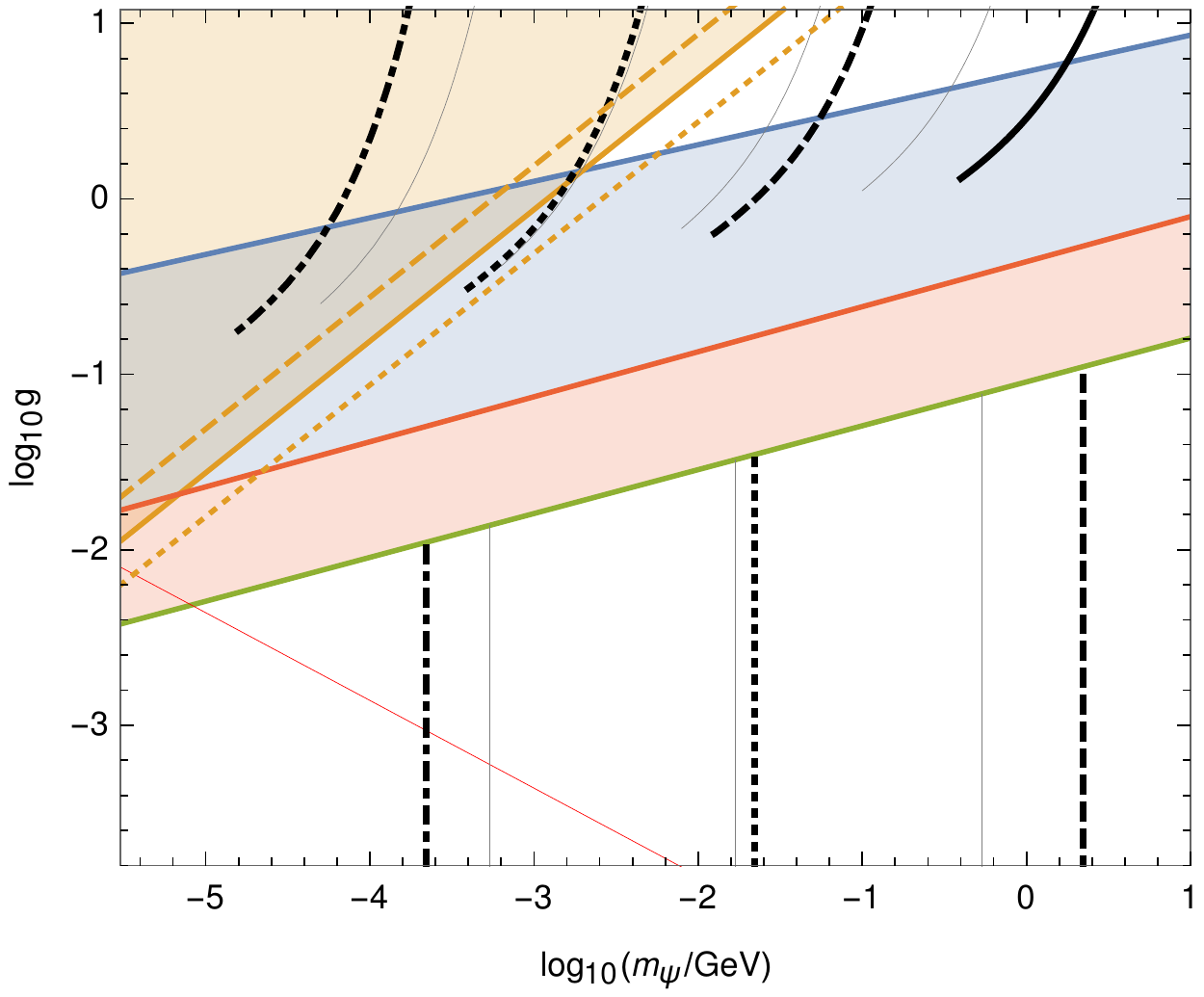}
\caption{Same as Figure \ref{fig2} for the case of fermion dark matter, for $m_{\rm s} = 10 m_\psi$. The red contour marks the thermalization via scalar self-scattering for $\lambda_{\rm s} = 0.1$ as in Figure \ref{fig: freeze in yield fermion}, and $\lambda_{\rm s} = 0.01$ has been used for the isocurvature contours.}
\label{fig5}
\end{center}
\end{figure}

A similar result can be derived for fermions. In the limit $g^2m_{\psi}/m_{\rm s}\ll 1$ the fermionic self-interaction cross-section divided by mass is 
\be
\label{psi_crosssec}
\frac{\sigma_{\psi}}{m_{\psi}} = \frac{g^4 m_\psi}{4\pi m_{\rm s}^4} .
\ee
If $\psi$ constitutes dominant fraction of total dark matter abundance, combining \eqref{isocurvature} and \eqref{psi_crosssec} gives an upper bound on $\psi$ mass
\be
\label{mpsi_bound}
\frac{m_{\psi}}{{\rm GeV}} \lesssim
\begin{cases}
5.9\times 10^4 g^{-4} (m_{\rm s}/{\rm GeV})^4 ,\\
3 \lambda_{\rm s}^{3/8}(H_*/10^{11} {\rm GeV})^{-3/2},
\end{cases}
\ee
where the former limit is given by the dark matter self-interaction limit \eqref{psi_crosssec} and the latter by the isocurvature bound \eqref{isocurvature}. The results are shown in Figure
\ref{fig5}.

\subsection{Other constraints}

Another potential constraint for light dark matter arises from the expansion rate of the universe during the Big Bang nucleosynthesis, typically formulated in terms of the effective number of neutrino species $N_\nu^{\rm eff} < 3.38$ \cite{Ade:2015xua}. However, as the entropy density of the dark sector in our scenario is always very small compared to the SM plasma, $\chi \gtrsim 10^3$ to produce the observed dark matter abundance, the dark sector never contributes significantly to the effective number of relativistic degrees of freedom, and this bound is always avoided in the otherwise allowed parameter space. 
A similar conclusion was also reached in \cite{Bernal:2015xba}.

As an extension of our model, we can consider a scenario where the singlet fermion, i.e. the sterile neutrino mixes with the active ones. Such a scenario is relevant, for example, as a dark matter interpretation of a spectral feature at $E\simeq 3.55$ keV observed in X-ray observations from several dark matter dominated sources \cite{Bulbul:2014sua,Boyarsky:2014jta}. While non-dark matter explanations of this observation have been suggested \cite{Riemer-Sorensen:2014yda,Jeltema:2014qfa,Carlson:2014lla}, the origin of the 3.5 keV line as dark matter decay has not been conclusively excluded \cite{Ruchayskiy:2015onc}. 

If dark matter is constituted by a sterile neutrino as in our model, and saturation of the observed abundance is assumed, then matching the intensity of the observed spectral line implies that the mixing angle should be $\sin^2(2\theta)\simeq 4.3\times 10^{-11}$ \cite{Bulbul:2014sua}. 
The energy of the observed line fixes the mass of the sterile neutrino to be $7.1$ keV. The small sterile-active mixing does not affect the stability of the dark matter candidate on the timescales relevant for our analysis of the relic abundance and associated constraints. Also, even if the non-resonant oscillations will contribute to the production of the dark matter abundace, this is expected to be of the order of few percent in the present case \cite{Bulbul:2014sua}. Hence, the results we have presented earlier remain valid also in the case of small mixing between the sterile and active neutrinos.

From Figure \ref{fig5}, we see that for the dark sector mass hierarchy $m_{\rm s} = 10 m_\psi$ as chosen here, the 7 keV fermion mass resides in the freeze-in region. Thus, within this mass hierarchy, the dark matter abundance can be produced via the standard freeze-in mechanism through the scalar portal, as long as the Yukawa coupling is below $g\lesssim 0.004$ and the scalar self-coupling is not too large (below $\lambda_{\rm s} \lesssim 10^{-2}$ for $g \gtrsim 10^{-4}$, or $\lambda_{\rm s} \lesssim 10^{-3}$ for $g \gtrsim 10^{-6}$), so that the dark sector does not thermalize. The correct relic abundance is then produced via freeze-in for the portal coupling $\lambda_{\rm hs}\simeq 4\times10^{-9}$. 

For this mass hierarchy, the self-interacting scenario $\sigma/m_\psi\sim 1 {\rm cm}^2/{\rm g}$ resides in the red region, where the observed dark matter abundance can not be produced. However, as the self-interaction cross-section (\ref{psi_crosssec}) is a function of the scalar mass, this situation changes for a different dark sector mass hierarchy. For $m_{\rm s}=100m_\psi$, the self-interacting case for \mbox{$m_\psi=7$ keV} corresponds to $g\simeq 0.2$, yielding a relativistic dark freeze-out of the fermion abundance. 

In conclusion, the 7 keV sterile neutrino scenario is compatible with either the freeze-in or dark freeze-out production mechanism of dark matter, depending on the mass spectrum of the singlet sector, and the self-interacting dark matter scenario can be achieved in the relativistic dark freeze-out case, given a large enough mass hierarchy within the dark sector.

\section{Conclusions}
\label{checkout}

In this paper we have considered observational consequences of self-interacting, initially non-thermal dark matter. We have considered a simple hidden sector model, where the dark matter particle is either a sterile neutrino coupled with the SM only via a scalar portal, or the portal scalar itself. The out-of-equilibrium production mechanism requires that the portal coupling is very small, ${\cal O}(10^{-10})$, and hastily one would conclude that such singlet sector would easily remain undetected.

As we have shown, despite the tiny portal coupling, there are observational consequences of this scenario. First and foremost, the inflationary fluctuations of the singlet scalar mediator, which does not thermalize with the SM, contain a significant isocurvature component which is imprinted on the dark matter abundance. To avoid these constraints, a lower limit must be imposed on the scalar self-coupling, which on the other hand is limited from above by astrophysical bounds on scalar dark matter self-interactions. Additionally, the non-negligible self-coupling may lead to thermalization within the dark sector, resulting in a dark freeze-out mechanism for the relic abundance, rather than the standard freeze-in scenario.

Together these consequences place stringent bounds on model parameters. We have found that if the dark matter abundance constitutes of $s$ particles, the scalar mass is bounded from above by the isocurvature bound and from below by its self-interactions, \eqref{ms_ubound}. In case the sterile neutrinos constitute the dark matter abundance, an upper bound can be derived, \eqref{mpsi_bound}. Furthermore, the self-interacting dark matter region, characterized by \mbox{$0.1\ {\rm cm}^2/{\rm g} \lesssim \sigma_{\rm DM}/m_{\rm DM}\lesssim 1 {\rm cm}^2/{\rm g}$}, is only accessible for very light dark matter masses, \mbox{$m_{\rm s} \lesssim 10\ {\rm keV}$} via freeze-in, and for \mbox{$m_{\rm s} \lesssim 0.1\ {\rm GeV}$} via dark freeze-out in the scalar dark matter scenario, see Figure \ref{fig2}. For sterile neutrino dark matter the self-interactions are suppressed by the mass of the scalar mediator, and the allowed range for self-interacting dark matter is even more constrained than in the scalar case, see Figure \ref{fig5}.

Finally, the model can be enlarged to allow for mixing between the sterile neutrino and the active ones. A tiny mixing allows for a dark matter interpretation of the spectral feature at $E \simeq 3.5$ keV observed in the X-ray spectra from several dark matter dominated sources.   

We have shown that even a simple isolated sector, containing only the portal scalar and a singlet fermion, can accommodate a multitude of thermal histories resulting in warm or cold dark matter, depending on the internal couplings within the dark sector. Work remains to be done to further investigate frozen-in dark sectors with even more structure, such as broken or unbroken gauge interactions and different mass hierarchies. The ongoing and planned CMB polarization experiments will probe the amplitude of the tensor perturbations down to $r\lesssim 10^{-3}$, corresponding to an inflationary scale $H_*\sim10^{13}$ GeV. As we have shown, a positive observation would immensely affect not only models of inflation, but also dark matter models of this kind, ruling out large portions of the parameter space.

\section*{Acknowledgements}
We thank K. Kainulainen, S. Nurmi, M. Raidal, C. Spethmann and H. Veerm\"ae for disucssions. This work has been supported by the Academy of Finland, \mbox{grant\# 267842}.
TT acknowledges financial support from the Research Foundation of the 
University of Helsinki. VV acknowledges financial support from Magnus 
Ehrnrooth foundation.

\bibliography{freezein_observations_err.bib}

\end{document}